\documentclass[12pt]{article}

\begin{document}

\title{The $0^{+}\rightarrow 0^{+}$ positron double-$\beta$ decay with emission of
two neutrinos in the nuclei $^{96}$Ru, $^{102}$Pd, $^{106}$Cd and $^{108}$Cd}
\author{P. K. Raina$^{1}$, A. Shukla$^{1}$, S. Singh$^{2}$, P. K. Rath$^{2}$ and J.
G. Hirsch$^{3}$ \\
$^{1}$Department of Physics and Meteorology, IIT Kharagpur-721302, India\\
$^{2}$Department of Physics, University of Lucknow, Lucknow-226007, India\\
$^{3}$Instituto de Ciencias Nucleares, Universidad Nacional Aut\'{o}noma de
M\'{e}xico,\\
A.P. 70-543 M\'{e}xico 04510 D.F., M\'exico}
\date{}
\maketitle

\begin{abstract}
Theoretical results for 
two neutrinos in the nuclei $^{96}$Ru, $^{102}$Pd, $^{106}$Cd and $^{108}$Cd
are presented. The study employs the Hartree-Fock-Bogoliubov model to obtain
the wave functions of the parent and daughter nuclei, in conjunction with
the summation method to estimate the double beta decay nuclear matrix
elements. The reliability of the intrinsic wave functions of $^{96,102}$Ru, $%
^{96}$Mo, $^{102,106,108}$Pd and $^{106,108}$Cd nuclei are tested by
comparing the theoretically calculated spectroscopic properties with the
available experimental data. Calculated half-lives $T_{1/2}^{2\nu }$ of $%
^{96}$Ru, $^{102}$Pd, $^{106}$Cd and $^{108}$Cd nuclei for 2$\nu $ $\beta
^{+}\beta ^{+}$, 2$\nu $ $\beta ^{+}EC$ and 2$\nu $ $ECEC$ modes are
presented. The effect of deformation on the nuclear transition matrix
element $M_{2\nu }$ is also studied. PACS Numbers: 23.40.Hc, 21.60.Jz,
23.20.-g, 27.60.+j\newline
\end{abstract}

\section{Introduction}

The two neutrino double beta (2$\nu $ $\beta \beta $) decay and the
neutrinoless double beta (0$\nu $ $\beta \beta $) decay can occur in four
different 
processes: double electron ($\beta ^{-}\beta ^{-}$) emission, double
positron $(\beta ^{+}\beta ^{+})$ emission, electron-positron conversion $%
(\beta ^{+}EC)$ and double electron capture $(ECEC)$. The later three
processes are energetically competing and we shall refer to them as positron
double beta decay (e$^{+}$DBD) modes. The 2$\nu $ $\beta ^{-}\beta ^{-}$
decay is allowed in the standard model of electroweak unification (SM) and
the half-life of this process has been already measured for about ten nuclei
out of 35 possible candidates. Hence, the absolute values of the nuclear
transition matrix elements (NTMEs) $M_{2\nu }$ can be extracted directly.
Consequently, the validity of different models employed for nuclear
structure calculations can be tested by calculating the $M_{2\nu }$. In case
of 2$\nu $ e$^{+}$DBD modes, experimental limits on half-lives have already
been given for 14 out of 34 possible isotopes. The observation of 2$\nu $ e$%
^{+}$DBD modes would further constrain the nuclear models employed to study
the $\beta \beta $ decay severely.

On the other hand, the 0$\nu $ $\beta \beta $ decay violates the lepton
number conservation and is possible in gauge theoretical models beyond the
SM 
as GUTs, Majoron models, R$_{p} $ violating SUSY models, lepto quark
exchange and compositeness scenario. The aim of all the present experimental
activities is to observe the 0$\nu $ $\beta \beta $ decay. The observation
of 0$\nu $ e$^{+}$DBD modes would play a crucial role in discriminating
finer issues like the dominance of Majorana neutrino mass or right handed
currents. The experimental aspects and theoretical implications of e$^{+}$%
DBD modes have been 
widely discussed over the past years [1-11].

The experimental study of $\beta ^{-}\beta ^{-}$ decay is usually preferable
due to a larger available phase space in comparison to e$^{+}$DBD modes.
However, the experimental sensitivity of $\beta ^{-}\beta ^{-}$ decay mode
gets limited because of the presence of electron background. On the other
hand, the e$^{+}$DBD modes are attractive from the experimental point of
view due to the fact that they can be easily separated from the background
contaminations and easily detected through coincidence signals from four $%
\gamma $-rays, two $\gamma $-rays and one $\gamma $-ray for $\beta ^{+}\beta
^{+}$, $\beta ^{+}EC$ and $ECEC$ modes respectively. In the case of the 2$%
\nu $ $ECEC$ mode, the \textit{Q}-value of $^{106}$Cd is pretty large, 
2.782 MeV, but the detection of the 0$^{+}\rightarrow $0$^{+}$ transition is
difficult since only X-rays are emitted.

In 1955, Winter 
studied the e$^{+}$DBD modes of $^{106}$Cd experimentally to explore the
possibility of distinguishing between the Dirac or Majorana character of the
electron neutrino \cite{win55}. The 2$\nu $ e$^{+}$DBD modes were studied
theoretically for the first time by Rosen and Primakoff \cite{ros65}.
Following the discovery of parity violation in beta decay, there was a
marked decline in the experimental searches of $\beta \beta $ decay in
general as both the lepton number conservation and the $\gamma _{5}$
invariance had to be violated for the 0$\nu $ $\beta \beta $ decay to occur.
However, the perception began to change after Vergados showed that e$^{+}$%
DBD modes are possible as lepton number violating process in gauge theories
beyond the SM \cite{ver83}. Kim and Kubodera estimated the half-lives of all
the three modes with modified NTMEs and non-relativistic phase space factors 
\cite{kim83}. Abad \textit{et al.} performed similar calculations using
relativistic Coulomb wave functions \cite{aba84}. Some other theoretical
studies were also done for the e$^{+}$DBD modes [15-18]. The experimental
activities on the study of 2$\nu $ e$^{+}$DBD modes were also resumed
[19-21]. In the meantime, the QRPA emerged as 
a successful model in explaining the quenching of NTMEs by incorporating the
particle-particle part of the effective nucleon-nucleon interaction in the
proton-neutron channel \cite{vog86} and the observed \textit{T}$_{1/2}^{2\nu
}$ of several 2$\nu $ $\beta ^{-}\beta ^{-}$ decay emitters were reproduced
successfully \cite{suh98}. Subsequently, the 2$\nu $ e$^{+}$DBD modes were
studied in shell model, QRPA and its extensions, SU(4)$_{\sigma \tau }$ and
SSDH and pseudo SU(3) \cite{suh98}.

Low-background setups using Ge detectors were proposed by Barabash \cite
{bar90} to detect the transition of 2$\nu $ $ECEC$ mode to the 0$_{1}^{+}$
excited state. New developments in experimental setups have led to good
limits on the measurement of the 2$\nu $ e$^{+}$DBD modes of nuclei of our
interest namely $^{106}$Cd [12,20,24--30] and $^{108}$Cd \cite
{geo95,kie03,dan03} through the direct counting experiments. In the mass
region A$\sim $100, Norman has studied the 2$\nu $ e$^{+}$DBD modes of $%
^{96} $Ru \cite{nor85} and $^{102}$Pd is also a potential candidate to be
studied with Q-value of about 1.175 MeV with natural abundance of about
1.02\%. With improved sensitivity in detection systems of the planned bigger
Osaka-OTO experiment \cite{ito97} and COBRA \cite{zub01}, it is expected
that 2$\nu $ e$^{+}$DBD modes will be in observable range in the near
future. Hence, a timely reliable prediction of the half-lives of $^{96}$Ru, $%
^{102}$Pd, $^{106}$Cd and $^{108}$Cd nuclei will be helpful in the ongoing
planning of future experimental setups.

The structure of nuclei in the mass region \textit{A}$\approx$100 is quite
complex. This mass region offers a nice example of shape transitions, i.e.
sudden onset of deformation at neutron number \textit{N}=60. The nuclei are
soft vibrators for \textit{N}$<$60 and quasi rotors for \textit{N}$>$ 60.
The nuclei with neutron number \textit{N}=60 are transitional nuclei. In
this mass region \textit{A}=96-108, the smallest and largest quadrupole
deformation parameter $\beta _{2}$ are 0.1580$\pm 0.0032$ and 0.2443$\pm
0.00 $30 for $^{96}$Ru and $^{102}$Ru respectively. Further, the pairing of
like nucleons plays an important role in all $\beta \beta $ decay emitters,
which are even-\textit{Z} and even-\textit{N} nuclei. Thus, it is expected
that pairing and deformation degrees of freedom will play some crucial role
in the structure of $^{96,102}$Ru, $^{96}$Mo, $^{102,106,108}$Pd and $%
^{106,108} $Cd nuclei. For the study of 2$\nu $ e$^{+}$DBD modes of these
nuclei, it is desirable to have a framework in which the pairing and
deformation degrees of freedom are treated on equal footing in its
formalism. The Projected Hartree-Fock-Bogoliubov (PHFB) model is, in this
sense, a sensible choice which fulfills these requirements. However in the
present version of the PHFB model, it is not possible to study the structure
of odd-odd nuclei. Hence, the single beta decay rates and the distribution
of Gamow-Teller strength can not be calculated. On the other hand, the study
of these processes has implications in the understanding of the role of the
isoscalar part of the proton-neutron interaction. This is a serious draw
back in the present formalism of the PHFB model. Notwithstanding, PHFB model
has been successfully applied to the $\beta^-\beta^-$ decay of many emitters
in this mass region, where it was possible to describe, in the same context,
the lowest excited states of the parent and daughter nuclei, as well as
their electromangetic transition strengths on one hand, and to reproduce
their measured double-beta decay rates on the other \cite{Chan05}.

The aim of nuclear many-body theory is to describe the observed properties
of nuclei in a coherent framework. The $\beta \beta $ decay can be studied
in the same framework as many other nuclear properties and decays.
Experimental studies involving in-beam $\gamma $-ray spectroscopy concerning
the level energies as well as electromagnetic properties has yielded a vast
amount of data over the past years. Although the availability of data
permits a rigorous and detailed critique of the ingredients of the
microscopic model that seeks to provide a description of nuclear $\beta
\beta $ decay, most of the calculations of 2$\nu $ e$^{+}$DBD transition
matrix elements performed so far but for the work of Barabash \textit{et al. 
}\cite{bar96} and Suhonen \textit{et al. }\cite{suh01} do not satisfy this
criterion. The successful study of 2$\nu $ e$^{+}$DBD modes of $^{106}$Cd
for $0^{+}\rightarrow 0^{+}$ transition together with other observed nuclear
properties, 
like the yrast spectra, reduced transition probabilities $B(E2$: $%
0^{+}\rightarrow 2^{+})$, static quadrupole moments $Q(2^{+})$ and $g$
-factors $g(2^{+})$ of both parent and daughter nuclei using the PHFB model
in conjunction with the summation method \cite{shu05} has motivated us to
apply the same framework to study the 2$\nu $ e$^{+}$DBD modes of $^{96}$Ru, 
$^{102}$Pd and $^{108}$Cd isotopes. The reason for 
presenting again the results of $^{106}$Cd 
is that the HFB\ wave functions are generated with improved accuracy and it
is nice to see that the results remain almost unchanged.

Further, it has been shown that there exists an inverse correlation between
the Gamow-Teller strength and quadrupole moment \cite{aue93, tro96}. The 
\textit{PPQQ} interaction \cite{bar68} has two 
terms, associated with the pairing interaction (\textit{PP}) and the
quadrupole-quadrupole (\textit{QQ}) interactions. The former accounts for
the sphericity of nucleus, whereas the later increases the collectivity in
the nuclear intrinsic wave functions and makes the nucleus deformed. Hence,
the PHFB model in conjunction with the \textit{PPQQ} interaction is a
convenient choice to examine the explicit role of deformation on the NTME\ $%
M_{2\nu }.$ In case of $^{106}$Cd, we have already shown that deformation
plays an important role in the variation of $M_{2\nu } $ vis-a-vis changing
strength of the \textit{QQ} part of effective two-body interaction \cite
{shu05}.

The structure of the present paper is as follows. The theoretical formalism
to calculate the half-lives of 2$\nu $ e$^{+}$DBD modes has been given in a
number of reviews \cite{doi92, suh98} and our earlier study of 2$\nu $ e$%
^{+} $DBD modes of $^{106}$Cd for the $0^{+}\rightarrow 0^{+}$ transition 
\cite{shu05}. Hence, we briefly outline steps of the above derivations in
sect. 2 for clarity in notation. Details of the mathematical expressions
used to calculate the spectroscopic properties of nuclei in the PHFB model
have been given by Dixit \textit{et al. }\cite{dix02}. In sect. 3, we
present results to check the reliability of the wave functions of $^{96,102}$%
Ru, $^{96}$Mo, $^{102,106,108}$Pd and $^{106,108}$Cd nuclei by calculating
the mentioned spectroscopic properties and 
comparing them with the available experimental data. The half-lives $%
T_{1/2}^{2\nu }$ for the 2$\nu $ e$^{+}$DBD modes of $^{96}$Ru, $^{102}$Pd, $%
^{106}$Cd and $^{108}$Cd nuclei for the 0$^{+}\rightarrow $0$^{+}$
transition are calculated. The role of deformation on $M_{2\nu }$ is also
studied. We present some concluding remarks in sect. 4.

\section{Theoretical framework}

The inverse half-life of the 2$\nu $ e$^{+}$DBD mode for the $%
0^{+}\rightarrow 0^{+}$ transition is given by

\begin{equation}
\left[ T_{1/2}^{2\nu }(0^{+}\rightarrow 0^{+})\right] ^{-1}=G_{2\nu }\left|
M_{2\nu }\right| ^{2} .
\end{equation}
The $G_{2\nu }$ is the integrated kinematical factor and the NTME $M_{2\nu }$
is expressed as 
\begin{equation}
M_{2\nu }=\sum\limits_{N}\frac{\langle 0_{F}^{+}||\mathbf{\sigma }\tau
^{-}||1_{N}^{+}\rangle \langle 1_{N}^{+}||\mathbf{\sigma }\tau
^{-}||0_{I}^{+}\rangle }{E_{0}+E_{N}-E_{I}} ,
\end{equation}
where $\ $%
\begin{equation}
E_{0}=\frac{1}{2}\left( E_{I}-E_{F}\right) =\frac{1}{2}W_{0} .
\end{equation}
The total energy released $W_{0}$ for different 2$\nu $ e$^{+}$DBD modes is
given by 
\begin{eqnarray}
W_{0}(\beta ^{+}\beta ^{+}) &=&Q_{\beta ^{+}\beta ^{+}}+2m_{e} , \\
W_{0}(\beta ^{+}EC) &=&Q_{\beta ^{+}EC}+e_{b} , \\
W_{0}(ECEC) &=&Q_{ECEC}-2m_{e}+e_{b1}+e_{b2} .
\end{eqnarray}

The summation over intermediate states is carried out using the summation
method \cite{civ93} and the NTME $M_{2\nu }$ can be written as 
\begin{equation}
M_{2\nu }=\frac{1}{E_{0}}\left\langle 0_{F}^{+}\left| \sum_{m}(-1)^{m}\Gamma
_{-m}F_{m}\right| 0_{I}^{+}\right\rangle ,  \label{eqcom}
\end{equation}
where the Gamow-Teller (GT) operator $\Gamma _{m}$ has been defined as

\begin{equation}
\Gamma _{m}=\sum_{s}\mathbf{\sigma }_{ms}\tau _{s}^{-} ,
\end{equation}
and 
\begin{equation}
F_{m}=\sum_{\lambda =0}^{\infty }\frac{(-1)^{\lambda }}{E_{0}^{\lambda }}%
D_{\lambda }\Gamma _{m} ,
\end{equation}
with 
\begin{equation}
D_{\lambda }\Gamma _{m}=\left[ H,\left[ H,........,\right. \left[ H,\Gamma
_{m}\right] .......\right] ^{(\lambda \hbox{~times})} .
\end{equation}
When the GT operator commutes with the effective two-body interaction, the
Eq. (\ref{eqcom}) can be further simplified to 
\begin{equation}
M_{2\nu }=\sum\limits_{\pi ,\nu }\frac{\langle 0_{F}^{+}||\mathbf{\sigma
.\sigma }\tau ^{-}\tau ^{-}||0_{I}^{+}\rangle }{E_{0}+\varepsilon (n_{\nu
},l_{\nu },j_{\nu })-\varepsilon (n_{\pi },l_{\pi },j_{\pi })} .  \label{m2n}
\end{equation}
The energy denominator is evaluated as follows. The difference in single
particle energies of neutrons in the intermediate nucleus and protons in the
parent nucleus is mainly due to the difference in Coulomb energies. Hence 
\begin{equation}
\varepsilon (n_{\nu },l_{\nu },j_{\nu })-\varepsilon (n_{\pi },l_{\pi
},j_{\pi })=\left\{ 
\begin{array}{llll}
\Delta _{C}-2E_{0} &  & for & n_{\nu }=n_{\pi },l_{\nu }=l_{\pi },j_{\nu
}=j_{\pi } \\ 
\Delta _{C}-2E_{0}+\Delta E_{s.o. spltting} &  & for & n_{\nu }=n_{\pi
},l_{\nu }=l_{\pi },j_{\nu }\neq j_{\pi }
\end{array}
\right. ,
\end{equation}
where the Coulomb energy difference $\Delta _{C}$ is given by Bohr and
Mottelson \cite{boh98} 
\begin{equation}
\Delta _{C}=\frac{0.70}{A^{1/3}}\left[ \left( 2Z+1\right) -0.76\left\{
\left( Z+1\right) ^{4/3}-Z^{4/3}\right\} \right] .
\end{equation}
In the case of pseudo SU(3) model [42-44], the energy denominator is a
well-defined quantity without any free parameter as the GT operator commutes
with the two-body interaction. The energy denominator has been evaluated
exactly for 2$\nu $ $\beta ^{-}\beta ^{-}$ \cite{cas94,hir95} and 2$\nu $ e$%
^{+}$DBD modes \cite{cer99} in pseudo SU(3) scheme.

In the present work, we use a Hamiltonian with \textit{PPQQ} type \cite
{bar68} of effective two-body interaction. Explicitly, the Hamiltonian is
written as 
\begin{equation}
{H}=H_{sp}+V(P)+\chi _{qq}V(QQ) ,  \label{hmtn}
\end{equation}
where $H_{sp}$ denotes the single particle Hamiltonian. The pairing part of
the effective two-body interaction $V(P)$ is written as 
\begin{equation}
V{(}P{)}=-\left( \frac{G}{4}\right) \sum\limits_{\alpha \beta
}(-1)^{j_{\alpha }+j_{\beta }-m_{\alpha }-m_{\beta }}a_{\alpha }^{\dagger
}a_{\bar{\alpha}}^{\dagger }a_{\bar{\beta}}a_{\beta } ,
\end{equation}
where $\alpha $ denotes the quantum numbers (\textit{nljm}). The state $\bar{%
\alpha}$ is same as $\alpha $ but with the sign of \textit{m} reversed. The 
\textit{QQ} part of the effective interaction $V(QQ)$\ is expressed as 
\begin{equation}
V(QQ)=-\left( \frac{\chi }{2}\right) \sum\limits_{\alpha \beta \gamma \delta
}\sum\limits_{\mu }(-1)^{\mu }\langle \alpha |q_{\mu }^{2}|\gamma \rangle
\langle \beta |q_{-\mu }^{2}|\delta \rangle \ a_{\alpha }^{\dagger }a_{\beta
}^{\dagger }\ a_{\delta }\ a_{\gamma } ,
\end{equation}
where 
\begin{equation}
{q_{\mu }^{2}}=\left( \frac{16\pi }{5}\right) ^{1/2}r^{2}Y_{\mu }^{2}(\theta
,\phi ) .
\end{equation}
The $\ \chi _{qq}$ is an arbitrary adimensional parameter and the final
results are obtained by setting the $\ \chi _{qq}$ = 1. The purpose of
introducing $\chi _{qq}$ is to study the role of deformation by varying the
strength of \textit{QQ} part of the effective two-body interaction.

The model Hamiltonian given by Eq. (\ref{hmtn}) is not isospin symmetric.
Hence, the energy denominator is not as simple as in Eq. (\ref{m2n}).
However, the violation of isospin symmetry for the \textit{QQ} part of our
model Hamiltonian is negligible, as will be evident from the parameters of
the two-body interaction given later. Further, the violation of isospin
symmetry for the pairing part of the two-body interaction is presumably
small in the mass region under study.

Under these assumptions, the expression to calculate the NTME $M_{2\nu }$ of
2$\nu $ e$^{+}$DBD modes for $0^{+}\rightarrow 0^{+}$ transition in the
PHFB\ model is obtained as follows.

In the PHFB model, states with good angular momentum $\mathbf{J}$ are
obtained from the axially symmetric HFB intrinsic state ${|\Phi _{0}\rangle }
$ with \textit{K}=0 using the standard projection technique \cite{oni66}
given by 
\begin{equation}
{|\Psi _{00}^{J}\rangle }=\left[ \frac{(2J+1)}{{8\pi ^{2}}}\right] \int
D_{00}^{J}(\Omega )R(\Omega )|\Phi _{0}\rangle d\Omega ,
\end{equation}
where $\ R(\Omega )$\ and $\ D_{00}^{J}(\Omega )$\ are the rotation operator
and the rotation matrix respectively. The axially symmetric HFB intrinsic
state ${|\Phi _{0}\rangle }$ can be written as 
\begin{equation}
{|\Phi _{0}\rangle }=\prod\limits_{im}(u_{im}+v_{im}b_{im}^{\dagger }b_{i%
\bar{m}}^{\dagger })|0\rangle ,
\end{equation}
where the creation operators $\ b_{im}^{\dagger }$\ and $\ b_{i\bar{m}%
}^{\dagger }$\ are defined as 
\begin{equation}
b{_{im}^{\dagger }}=\sum\limits_{\alpha }C_{i\alpha ,m}a_{\alpha m}^{\dagger
}\quad \hbox{and}\mathrm{\quad }b_{i\bar{m}}^{\dagger }=\sum\limits_{\alpha
}(-1)^{l+j-m}C_{i\alpha ,m}a_{\alpha ,-m}^{\dagger } .
\end{equation}
The results of HFB calculations are summarized by the amplitudes $%
(u_{im},v_{im})$ and expansion coefficients $C_{ij,m}$.

Finally, one obtains the following expression for NTME $M_{2\nu }$ of the 2$%
\nu $ e$^{+}$DBD mode:

\begin{eqnarray}
M_{2\nu } &=&\sum\limits_{\pi ,\nu }\frac{\langle {\Psi _{00}^{J_{f}=0}}|| 
\mathbf{\sigma .\sigma }\tau ^{-}\tau ^{-}||{\Psi _{00}^{J_{i}=0}}\rangle }{
E_{0}+\varepsilon (n_{\nu },l_{\nu },j_{\nu })-\varepsilon (n_{\pi },l_{\pi
},j_{\pi })}  \nonumber \\
&=&[n_{Z-2,N+2}^{J_{f}=0}n_{Z,N}^{J_{i}=0}]^{-1/2}\int\limits_{0}^{\pi
}n_{(Z,N),(Z-2,N+2)}(\theta )\sum\limits_{\alpha \beta \gamma \delta }\frac{%
\left\langle \alpha \beta \left| \mathbf{\sigma }_{1}.\mathbf{\sigma }%
_{2}\tau ^{-}\tau ^{-}\right| \gamma \delta \right\rangle }{%
E_{0}+\varepsilon _{\alpha }(n_{\nu },l_{\nu },j_{\nu })-\varepsilon
_{\gamma }(n_{\pi },l_{\pi },j_{\pi })}  \nonumber \\
&&\times \sum_{\varepsilon \eta }\frac{(f_{Z-2,N+2}^{(\nu )*})_{\varepsilon
\beta }}{\left[ 1+F_{Z,N}^{(\nu )}(\theta )f_{Z-2,N+2}^{(\nu )*}\right]
_{\varepsilon \alpha }^{{}}}\frac{(F_{Z,N}^{(\pi )*})_{\eta \delta }}{\left[
1+F_{Z,N}^{(\pi )}(\theta )f_{Z-2,N+2}^{(\pi )*}\right] _{\gamma \eta }^{{}}}
\sin \theta d\theta ,  \label{eqf}
\end{eqnarray}
where

\begin{equation}
n^{J}=\int\limits_{0}^{\pi }\{\det [1+F^{(\pi )}(\theta )f^{(\pi )\dagger
}]\}^{1/2}\times \{\det [1+F^{(\nu )}(\theta )f^{(\nu )\dagger
}]\}^{1/2}d_{00}^{J}(\theta )\sin (\theta )d\theta ,
\end{equation}
and

\begin{equation}
n_{(Z,N),(Z-2,N+2)}(\theta )=\{\det [1+F_{Z,N}^{(\pi )}(\theta
)f_{Z-2,N+2}^{(\pi )\dagger }]\}^{1/2}\times \{\det [1+F_{Z,N}^{(\nu
)}(\theta )f_{Z-2,N+2}^{(\nu )\dagger }]\}^{1/2} .
\end{equation}
The $\pi (\nu )$\ represents the proton (neutron) of nuclei involved in the 2%
$\nu $ e$^{+}$DBD. The matrices $f_{Z,N}$\ \ and $F_{Z,N}(\theta )\ $are
given by

\begin{equation}
\lbrack f_{Z,N}]_{\alpha \beta }=\sum_{i}C_{ij_{\alpha },m_{\alpha
}}C_{ij_{\beta },m_{\beta }}\left( v_{im_{\alpha }}/u_{im_{\alpha }}\right)
\delta _{m_{\alpha },-m_{\beta }} ,
\end{equation}
and 
\begin{equation}
\lbrack F_{Z,N}(\theta )]_{\alpha \beta }=\sum_{m_{\alpha }^{^{\prime
}}m_{\beta }^{^{\prime }}}d_{m_{\alpha },m_{\alpha }^{^{\prime
}}}^{j_{\alpha }}(\theta )d_{m_{\beta },m_{\beta }^{^{\prime }}}^{j_{\beta
}}(\theta )f_{j_{\alpha }m_{\alpha }^{^{\prime }},j_{\beta }m_{\beta
}^{^{\prime }}} .  \label{eq2}
\end{equation}

The calculation of NTME $M_{2\nu }^{{}}$ for the 2$\nu $ e$^{+}$DBD mode is
carried on as follows. In the first step, the matrices $[f_{Z,N}]_{\alpha
\beta }$ and $[F_{Z,N}(\theta )]_{\alpha \beta }$ are set up using
expressions given by Eqs. (24) and (25) respectively. Finally, the required
NTME $M_{2\nu }$ is calculated in a straight forward manner using Eq. (\ref
{eqf}) with 20 gaussian quadrature points in the range (0, $\pi $).

\section{Results\ and discussions}

The model space, single particle energies (SPE's) and the effective two-body
interaction are 
the same employed in our earlier calculation on 2$\nu $ e$^{+}$DBD modes of $%
^{106}$Cd for the $0^{+}\rightarrow 0^{+}$ transition \cite{shu05}. However,
we present a brief discussion of them in the following for convenience. The
model space consists of 1\textit{p}$_{1/2},$ 2\textit{s}$_{1/2,}$ 1\textit{d}%
$_{3/2}$, 1\textit{d}$_{5/2}$, 0\textit{g}$_{7/2}$, 0\textit{g}$_{9/2}$ and 0%
\textit{h}$_{11/2}$ orbits for protons and neutrons, where we have treated
the doubly even nucleus $^{76}$Sr (\textit{N}=\textit{Z}=38) as an inert
core. The orbit 1\textit{p}$_{1/2}$ has been included in the valence space
to examine the role of the \textit{Z}=40 proton core vis-a-vis the onset of
deformation in 
highly neutron rich isotopes. The set of single particle energies (SPE's)
used here are (in MeV) $\varepsilon $(1\textit{p}$_{1/2}$)=-0.8, $%
\varepsilon $(0\textit{g}$_{9/2}$)=0.0, $\varepsilon $(1\textit{d}$_{5/2}$%
)=5.4, $\varepsilon $(2\textit{s}$_{1/2}$)=6.4, $\varepsilon $(1\textit{d}$%
_{3/2}$)=7.9, $\varepsilon $(0\textit{g}$_{7/2}$)=8.4 and $\varepsilon $(0%
\textit{h}$_{11/2}$)=8.6 for proton and neutrons. This set of SPE's but for
the $\varepsilon $(0\textit{h}$_{11/2}$ ), which is slightly lowered, has
been employed in a number of successful shell model \cite{ver71} as well as
variational model \cite{kho82} calculations for nuclear properties in the
mass region \textit{A}$\approx$ 100.

The strengths of the pairing interaction has been fixed through the relation 
$G_{p}$ =30/\textit{A} MeV and $G_{n}$=20/\textit{A} MeV, which are same as
used by Heestand \textit{et al.} \cite{hee69} to explain the experimental $%
g(2^{+})$ data of some even-even Ge, Se, Mo, Ru, Pd, Cd and Te isotopes in
Greiner's collective model \cite{gre66}. The strengths of the like particle
components of the $QQ$ interaction are taken as: $\chi _{pp}$ = $\chi _{nn}$
= 0.0105 MeV \textit{b}$^{-4}$, where \textit{b} is oscillator parameter.
The strength of proton-neutron (\textit{pn}) component of the $QQ$
interaction $\chi _{pn}$ is varied 
to fit the spectra of 
$^{96,102}$Ru, $^{96}$Mo, $^{102,106,108}$Pd and $^{106,108}$Cd in 
agreement with the experimental results. To be more specific, we have taken
the theoretical spectra to be the optimum one if the excitation energy of
the $\ $2$^{+}$ state \ $E_{2^{+}}$ is reproduced as closely as possible to
the experimental value. Thus for a given model space, SPE's, $G_{p}$, $G_{n}$
and $\chi _{pp}$, we have fixed $\chi _{pn}$ through the experimentally
available energy spectra. We have given the values of $\chi _{pn}$ in Table
1. These values for the strength of the $QQ$ interaction are comparable to
those suggested by Arima on the basis of an empirical analysis of the
effective two-body interactions \cite{ari81}. All the parameters are kept
fixed throughout the calculation.

\subsection{The yrast spectra and electromagnetic properties}

In table 1, we have displayed the theoretically calculated and
experimentally observed values of yrast spectra for J$^{\pi }=$2$^{+},$ 4$%
^{+}$and 6$^{+}$ states of $^{96,102}$Ru, $^{96}$Mo, $^{102,106,108}$Pd and $%
^{106,108}$Cd isotopes. 
The agreement between the experimentally observed \ \cite{sak84} and
theoretically reproduced $E_{2^{+}}$ is quite good. However, it 
can be noticed that the theoretical spectra is more expanded in comparison
to the experimental spectra. This can be corrected to some extent in the
PHFB model in conjunction with the VAP prescription \cite{kho82}. However,
our aim is to reproduce properties of the low-lying 2$^{+}$ state. Hence, we
have not attempted to invoke the VAP prescription, which will unnecessarily
complicate the calculations.

In table 2, we present the calculated as well as the experimentally observed
values of the reduced $\ B(E2$:$0^{+}\to 2^{+})$ transition probabilities 
\cite{ram87}, static quadrupole moments $\ Q(2^{+})$ and the gyromagnetic
factors $\ g(2^{+})$ \cite{rag89}. In case of $B(E2$:$0^{+}\to 2^{+})$, only
some representative experimental values are tabulated. $B(E2$:$0^{+}\to
2^{+})$ results are given for effective charges $\ e_{eff}$ =0.40, 0.50 and
0.60 in columns 2 to 4, respectively. The experimentally observed values are
displayed in column 5. The calculated values are in excellent agreement with
the observed $B(E2$:$0^{+}\to 2^{+})$ of all the nuclei considered at $\
e_{eff}$ =0.50 except for $^{102}$Ru, $^{102}$Pd and $^{108}$Pd, which
differ by 0.049, 0.02 and 0.046 \textit{e}$^{2}$ b$^{2}$ respectively from
the experimental lower limits.

The theoretically calculated $\ Q(2^{+})$ values are tabulated in columns 6
to 8 of the same table 2, along with the experimentally observed $Q(2^{+})$
data in column 9, for the same effective charges as used in case of $B(E2$:$%
0^{+}\to 2^{+})$. Again, the agreement between the calculated and
experimental $Q(2^{+})$ values is quite good in case of $^{102}$Ru, $^{108}$%
Cd, $^{106}$Pd and $^{108}$Pd nuclei except for $^{106}$Cd where the
difference is 0.2 \textit{e} b. In case of $^{96}$Ru,$^{96}$Mo and $^{102}$%
Pd, although the experimental values have large error bars and a meaningful
comparison is difficult, the agreement between calculated and observed
values is not satisfactory. The \ $g(2^{+})$ values are calculated with $\
g_{l}^{\pi }$=1.0, $\ g_{l}^{\nu }$=0.0, $\ g_{s}^{\pi }$=$\ g_{s}^{\nu }$
=0.60. No experimental result for $g(2^{+})$ is available for the isotope $%
^{96}$Ru. The calculated and experimentally observed$\ g(2^{+})$ values are
in excellent agreement for $^{102}$Ru, $^{102}$Pd, $^{106}$Cd and $^{108} $%
Cd nuclei whereas they are off by 0.073, 0.046 and 0.093 nm only for $^{96}$%
Mo, $^{106}$Pd and $^{108}$Pd isotopes, respectively.

From the above discussions, it is clear that the overall agreement between
the calculated and observed electromagnetic properties is quite good. Hence,
the PHFB wave functions of $^{96,102}$Ru, $^{96}$Mo, $^{102,106,108}$Pd and $%
^{106,108}$Cd nuclei generated by fixing $\chi _{pn}$ to reproduce the yrast
spectra are quite reliable. Below, we present the results of NTMEs $M_{2\nu
} $ as well as the half-lives $T_{1/2}^{2\nu }$ of $^{96}$Ru, $^{102}$Pd,$%
^{106}$Cd and $^{108}$Cd for the$\ 0^{+}\to 0^{+}$ transition using the same
HFB wavefunctions.

\subsection{Results of $\ 2\nu $ $\beta ^{+}\beta ^{+}/\beta ^{+}EC/ECEC$
decay}

In table 3, we have compiled the available experimental and theoretical
results for 2$\nu $ e$^{+}$DBD modes of $^{96}$Ru, $^{102}$Pd,$^{106}$Cd and 
$^{108}$Cd nuclei along with our calculated NTMEs \textit{M}$_{2\nu }$ and
the corresponding half-lives $T_{1/2}^{2\nu }$. The calculated phase space
factors were obtained following the prescription of Doi \textit{et al. }\cite
{doi92} in the approximation $C_{1}=1.0,$ $C_{2}=0.0$, $C_{3}=0.0$ and $%
R_{1,1}(\varepsilon )=R_{+1}(\varepsilon )+R_{-1}(\varepsilon )=$1.0. The
phase space integrals have been evaluated for $\ g_{A}$= 1.261 by Doi 
\textit{et al.} \cite{doi92}. However, in heavy nuclei it is more justified
to use the nuclear matter value of $\ g_{A}$ around 1.0. Hence, the
theoretical $T_{1/2}^{2\nu }$ are 
presented both for $\ g_{A}$=1.0 and 1.261.

In case of $^{96}$Ru, the half-life limits $T_{1/2}^{2\nu }$ of the 2$\nu $ $%
\beta ^{+}EC$ and 2$\nu $ $ECEC$ modes for the $0^{+}\rightarrow 0^{+}$
transition has been investigated by Norman \cite{nor85} and are of the order
of 10$^{16}$ y. The calculated NTMEs $M_{2\nu }$ in PHFB and SU(4)$_{\sigma
\tau }$ \cite{rum98} model differ by a factor of 2 for all the three modes
while in QRPA model \cite{hir94}, the values of NTMEs $M_{2\nu }$ are larger
than the PHFB\ model values by a factor of 5, approximately. The phase space
factors for $^{96}$Ru isotope are $G_{2\nu }(\beta ^{+}\beta ^{+})$ = 2.516$%
\times $10$^{-26}$ y$^{-1}$, $G_{2\nu }(\beta ^{+}EC)$ = 9.635$\times $ 10$%
^{-22}$ y$^{-1}$ and $G_{2\nu }(ECEC)$ = 5.385$\times $10$^{-21}$ y$^{-1}$.
The theoretically calculated \textit{T}$_{1/2}^{2\nu }$ are of the order of
10$^{26-28}$ y, 10$^{22-23}$ y and 10$^{21-23}$ y for 2$\nu $ $\beta
^{+}\beta ^{+}$, 2$\nu $ $\beta ^{+}EC$ and 2$\nu $ $ECEC$ modes
respectively for $g_{A}$ = 1.261-1.00.

The e$^{+}$DBD modes of $^{102}$Pd isotope for $0^{+}\rightarrow 0^{+}$
transition has been investigated neither experimentally nor theoretically so
far. We have used the phase space factors $G_{2\nu }(\beta ^{+}EC)$ = 1.449$%
\times $10$^{-30}$ y$^{-1}$ and $G_{2\nu }(ECEC)$ = 9.611$\times $ 10$^{-23}$
y$^{-1}$ for 2$\nu $ $\beta ^{+}EC$ and 2$\nu $ $ECEC$ modes, respectively.
In PHFB model, the predicted $T_{1/2}^{2\nu }$ of 2$\nu $ $\beta ^{+}EC$ and
2$\nu $ $ECEC$ modes are (2.509-6.344)$\times $10$^{32}$ y and (3.783-9.565)$%
\times $10$^{24}$ y respectively for $g_{A}$ = 1.261-1.00.

We have compiled the available experimental [12,20,24-30] and theoretical
results [26,34,54-60] for $^{106}$Cd along with our calculated $M_{2\nu }$
and corresponding half-life $T_{1/2}^{2\nu }$ in table 3. In the case of $%
^{106}$Cd, the phase factors are $G_{2\nu }(\beta ^{+}\beta ^{+})$ = 4.263$%
\times $10$^{-26}$ y$^{-1}$, $G_{2\nu }(\beta ^{+}EC)\ $= 1.570$\times $10$%
^{-21}$ y$^{-1}$and $G_{2\nu }(ECEC)$ = 1.152$\times $ 10$^{-20}$ y$^{-1}$,
respectively. In comparison to the theoretically predicted $T_{1/2}^{2\nu },$
the present experimental limits for 0$^{+}\rightarrow $0$^{+}$ transition of 
$^{106}$Cd are smaller by a factor of 10$^{5-7}$ in case of 2$\nu $ $\beta
^{+}\beta ^{+}$ mode but are quite close for 2$\nu $ $\beta ^{+}EC$ and 2$%
\nu $ $ECEC$ modes. The half-life $T_{1/2}^{2\nu }$ calculated in PHFB model
using the summation method differs from all the existing calculations. The
presently calculated NTME $M_{2\nu }$ is smaller than the recently given
results in QRPA(WS) model of Suhonen and Civitarese \cite{suh01} by a factor
of 2 approximately for all the three modes. The theoretical $M_{2\nu }$
values of PHFB model and SU(4)$_{\sigma \tau }$ \cite{rum98} again differ by
a factor of 2 approximately for the 2$\nu $ $\beta ^{+}EC$ and 2$\nu $ $ECEC$
modes. On the other hand, the $M_{2\nu }$ calculated in our PHFB model is
smaller than the values of Hirsch \textit{et al.} \cite{hir94} by a factor
of 3 approximately in case of 2$\nu $ $\beta ^{+}\beta ^{+}$ and 2$\nu $ $%
ECEC$ modes while for 2$\nu $ $\beta ^{+}EC$ mode the results differ by a
factor of 4 approximately. All the rest of the calculations predict NTMEs,
which are larger than our predicted $M_{2\nu }$ approximately by a factor of
7 \cite{suh93,toi97} to 10 \cite{bar96}. The predicted $T_{1/2}^{2\nu }$ of 2%
$\nu $ $\beta ^{+}\beta ^{+}$, 2$\nu $ $\beta ^{+}EC$ and 2$\nu $ $ECEC$
modes in PHFB model are $(3.495-8.836)\times $10$^{27}$ y, $%
(9.489-23.992)\times $10$^{22}$ y and $(1.293-3.270)\times $10$^{22}$ y
respectively for $g_{A}$=1.26 and 1.0.

The 2$\nu $ $ECEC$ mode of $^{108}$Cd for the $0^{+}\rightarrow 0^{+}$
transition has been investigated by Georgadze \textit{et al. }\cite{geo95}%
\textit{, }Kiel\textit{\ et al. }\cite{kie03}\textit{\ }and\textit{\ }%
Danevich\textit{\ et al. } \cite{dan03}\textit{.} No theoretical calculation
has been done so far to study the above mentioned mode of $^{108}$Cd
isotope. The phase space factor of 2$\nu $ $ECEC$ mode is $G_{2\nu }(ECEC)$
= 2.803$\times $10$^{-26}$ y$^{-1}$. In PHFB model, the calculated half-life 
$T_{1/2}^{2\nu }$ of the 2$\nu $ $ECEC$ decay mode is $3.939\times $10$^{27}$
y and $9.959\times $10$^{27}$ y for $g_{A}$=1.261 and 1.0 respectively.

The quenching of the nuclear matrix elements seems to be closely related
with the explicit inclusion of deformation effects, which are absent in the
other models. We analyze in detail this point below.

\subsection{Deformation effect}

We have investigated the variation of the $\left\langle
Q_{0}^{2}\right\rangle $, $\beta _{2}$ and $M_{2\nu }$ with respect to the
change in strength of the \textit{QQ} interaction $\chi _{qq}$ to understand
the role of deformation on the NTME $M_{2\nu }$. Out of several
possibilities, we have taken the quadrupole moment of the intrinsic state $%
\left\langle Q_{0}^{2}\right\rangle $ (in arbitrary units) and the
quadrupole deformation parameter $\beta _{2}$ as a quantitative measure of
the deformation. The quadrupole moment of the intrinsic states $\left\langle
Q_{0}^{2}\right\rangle $, deformation parameter $\beta _{2}$ and the NTMEs $%
M_{2\nu }$ for different $\chi _{qq}$ are tabulated in table 4. The
deformation parameter has been calculated with the same effective charge as
used in the calculation of $B(E2$:$0^{+}\to 2^{+})$ transition probabilities.

It is noticed that the $\left\langle Q_{0}^{2}\right\rangle $ as well as $%
\beta _{2}$ increases in general as the $\chi _{qq}$ is varied from 0 to 1.5
except a few anomalies. The intrinsic quadrupole moments show fluctuations
in case of $^{96}$Ru at $\chi _{qq}$ value 0.6. In case of $^{96}$Mo,
similar fluctuations are observed at $\chi _{qq}$ equal to 0.4 and 0.6. In
case of $^{102,106}$Pd nuclei, the fluctuations occur at $\chi _{qq}$=0.4.
In all cases, it is found that the quadrupole deformation parameter $\beta
_{2}$ follows the same behavior as the quadrupole moment of the intrinsic
state $\left\langle Q_{0}^{2}\right\rangle $ with respect to the change in $%
\chi _{qq}$ except for $^{106}$Cd isotope. In this case, the $\left\langle
Q_{0}^{2}\right\rangle $ increases but $\beta _{2}$ decreases at $\chi _{qq}$%
=0.4. Further, there is an anticorrelation between the deformation parameter
and the NTME $M_{2\nu }$ in general but for a few exceptions.

To quantify the effect of deformation on $M_{2\nu }$, we define a quantity $%
D_{2\nu }$ as the ratio of $M_{2\nu }$ at zero deformation ($\chi _{qq}=0$)
and full deformation ($\chi _{qq}=1$). The $D_{2\nu }$ is given by 
\begin{equation}
D_{2\nu }=\frac{M_{2\nu }(\chi _{qq}=0)}{M_{2\nu }(\chi _{qq}=1)} .
\end{equation}
The values of $D_{2\nu }$ are 3.13, 3.40, 2.06 and 2.19 for $^{96}$Ru,$%
^{102} $Pd, $^{106}$Cd and $^{108}$Cd nuclei respectively. These values of $%
D_{2\nu }$ suggest that the $M_{2\nu }$ is quenched by a factor of 2 to 3.5
approximately in the mass region 96$\leq A\leq $108 due to deformation
effects.

Given the schematic nature of the PPQQ interaction employed in the present
calculation, and the fact that many of the nuclei studied are in the
transtional region and do not display a well defined rotational spectrum,
the quenching factors discussed above could be considered as a conservative
estimate of the uncertainties in the predicted $2\nu$ nuclear matrix
elements. They qualify both the present results and those obtained with
other models where deformation is not explicitely considered. The
uncertainties associated with the $0\nu$ processes would be expected to be
far smaller. Future work is expected to clarify this point. 

\section{Conclusions}

To summarize, we have tested the quality of PHFB wave functions by comparing
the theoretically calculated results for yrast spectra, reduced $B(E2$:$%
0^{+}\rightarrow 2^{+})$ transition probabilities, static quadrupole moments 
$Q(2^{+})$ and $g$-factors $g(2^{+})$ of $^{96,102}$Ru, $^{96}$Mo, $%
^{102,106,108}$Pd and $^{106,108}$Cd nuclei participating in 2$\nu $ e$^{+}$
DBD modes with the available experimental results. The same PHFB wave
functions are employed to calculate NTMEs $M_{2\nu }$ and half-lives \textit{%
T}$_{1/2}^{2\nu }$ of $^{96}$Ru (2$\nu $ $\beta ^{+}\beta ^{+}$, 2$\nu $ $%
\beta ^{+}EC$ and 2$\nu $ $ECEC$ modes), $^{102}$Pd (2$\nu $ $\beta ^{+}EC$
and 2$\nu $ $ECEC$ modes), $^{106}$Cd (2$\nu $ $\beta ^{+}\beta ^{+}$, 2$\nu 
$ $\beta ^{+}EC$ and 2$\nu $ $ECEC$ modes) and $^{108}$Cd (2$\nu $ $ECEC$
mode) nuclei. It is noticed that the proton-neutron part of the \textit{PPQQ}
interaction, which is responsible for triggering deformation in the
intrinsic ground state, plays an important role in the quenching of \textit{M%
}$_{2\nu }$ by a factor of approximately 2 to 3.5 in the considered mass
region 96$\leq A\leq $108. In case of $^{96}$Ru and $^{106}$Cd, we have
presented and discussed the theoretical results of 2$\nu $ e$^{+}$DBD modes
in the PHFB model along with other available nuclear models for $%
0^{+}\rightarrow 0^{+}$ transition . In case of $^{102}$Pd and $^{108}$Cd,
These are the first theoretical calculations and in view of growing
interests in the study of 2$\nu $ e$^{+}$DBD modes, these predictions would
be helpful in the planning of future experimental setups.

{\small \noindent }\noindent \noindent Table 1. Excitation energies (in MeV)
of J$^{\pi }=$2$^{+},$ 4$^{+}$ and 6$^{+}$ yrast states of $^{96,102}$Ru, $%
^{96}$Mo, $^{102,106,108}$Pd and $^{106,108}$Cd nuclei. 

{\small \bigskip }

\noindent 
\begin{tabular}{llllllllll}
\hline\hline
Nucleus & $\chi _{pn}$ &  & Theo. & Expt. \cite{sak84} & Nucleus & $\chi
_{pn}$ &  & Theo. & Expt. \cite{sak84} \\ \hline
$_{44}^{96}$Ru & 0.02417 & $E_{2^{+}}$ & 0.8323 & 0.8326 & $_{42}^{96}$Mo & 
0.02557 & $E_{2^{+}}$ & 0.7779 & 0.7782 \\ 
&  & $E_{4^{+}}$ & 2.1389 & 1.51797 &  &  & $E_{4^{+}}$ & 2.0373 & 1.6282 \\ 
&  & $E_{6^{+}}$ & 3.8037 & 2.1496 &  &  & $E_{6^{+}}$ & 3.5775 & 2.4406 \\ 
\hline
$_{46}^{102}$Pd & 0.01573 & $E_{2^{+}}$ & 0.5551 & 0.5565 & $_{44}^{102}$Ru
& 0.02054 & $E_{2^{+}}$ & 0.4751 & 0.4751 \\ 
&  & $E_{4^{+}}$ & 1.6010 & 1.2760 &  &  & $E_{4^{+}}$ & 1.4773 & 1.1064 \\ 
&  & $E_{6^{+}}$ & 2.9467 & 2.1115 &  &  & $E_{6^{+}}$ & 2.8737 & 1.8732 \\ 
\hline
$_{48}^{106}$Cd & 0.01505 & $E_{2^{+}}$ & 0.6321 & 0.6327 & $_{46}^{106}$Pd
& 0.01441 & $E_{2^{+}}$ & 0.5115 & 0.5119 \\ 
&  & $E_{4^{+}}$ & 1.7298 & 1.4939 &  &  & $E_{4^{+}}$ & 1.4816 & 1.2292 \\ 
&  & $E_{6^{+}}$ & 3.1610 & 2.4918 &  &  & $E_{6^{+}}$ & 2.7264 & 2.0766 \\ 
\hline
$_{48}^{108}$Cd & 0.01481 & $E_{2^{+}}$ & 0.6319 & 0.6330 & $_{46}^{108}$Pd
& 0.01443 & $E_{2^{+}}$ & 0.4336 & 0.4339 \\ 
&  & $E_{4^{+}}$ & 1.8072 & 1.5084 &  &  & $E_{4^{+}}$ & 1.3126 & 1.0482 \\ 
&  & $E_{6^{+}}$ & 3.3138 & 2.5413 &  &  & $E_{6^{+}}$ & 2.4826 & 1.7712 \\ 
\hline\hline
\end{tabular}

\vskip .9cm 

\noindent Table 2. Comparison of the calculated and experimentally observed
reduced transition probability $B(E2$:$0^{+}\rightarrow 2^{+})$ in e$^{2}$ b$%
^{2}$, static quadrupole moments $Q(2^{+})$ in e b and $g$-factors $g(2^{+})$
in nuclear magneton. Here $B(E2)$ and $Q(2^{+})$ are calculated for
effective charge $e_{p}=$1+$e_{eff}$ and $e_{n}=e_{eff}$. The $g(2^{+})$ has
been calculated for $g_{l}^{\pi }=$1.0, $g_{l}^{\nu }=$0.0 and $g_{s}^{\pi
}=g_{s}^{\nu }=$0.60. 

\bigskip 

\noindent 
\begin{tabular}{ccccccccccc}
\hline\hline
Nuclei & \multicolumn{4}{c}{$B(E2$:$0^{+}\rightarrow 2^{+})$} & 
\multicolumn{4}{c}{$Q(2^{+})$} & \multicolumn{2}{c}{$g(2^{+})$} \\ 
& \multicolumn{3}{c}{Theo.} & Expt. \cite{ram87} & \multicolumn{3}{c}{Theo.}
& Expt. \cite{rag89} & Theo. & Expt. \cite{rag89} \\ \cline{2-4}\cline{6-8}
&  & $e_{eff}$ &  &  &  & $e_{eff}$ &  &  &  &  \\ \cline{2-4}\cline{6-8}
& 0.40 & 0.50 & 0.60 &  & 0.40 & 0.50 & 0.60 &  &  &  \\ \hline
$^{96}$Ru & 0.208 & 0.261 & 0.319 & 0.251$\pm 0.010$ & -0.412 & -0.461 & 
-0.510 & -0.15$\pm 0.27$ & 0.394 &  \\ 
&  &  &  & 0.260$\pm 0.010$ &  &  &  &  &  &  \\ 
$^{96}$Mo & 0.265 & 0.335 & 0.413 & 0.310$\pm 0.047$ & -0.466 & -0.524 & 
-0.582 & -0.20$\pm 0.08$ & 0.563 & 0.419$\pm 0.033\pm 0.038^{*}$ \\ 
&  &  &  & 0.271$\pm 0.005$ &  &  &  & +0.04$\pm 0.08$ &  &  \\ \hline
$^{102}$Pd & 0.323 & 0.410 & 0.507 & 0.460$\pm 0.030$ & -0.514 & -0.580 & 
-0.645 & -0.20$\pm 0.20$ & 0.386 & 0.41$\pm 0.04$ \\ 
&  &  &  &  &  &  &  &  &  & 0.39$\pm 0.05$ \\ 
$^{102}$Ru & 0.458 & 0.585 & 0.726 & 0.640$\pm 0.006$ & -0.613 & -0.692 & 
-0.771 & -0.57$\pm 0.07$ & 0.385 & 0.371$\pm 0.031$ \\ 
&  &  &  & 0.651$\pm 0.016$ &  &  &  & -0.68$\pm 0.08$ &  &  \\ \hline
$^{106}$Cd & 0.330 & 0.422 & 0.525 & 0.410$\pm 0.$02$0$ & -0.518 & -0.586 & 
-0.654 & -0.28$\pm 0.08$ & 0.372 & 0.40$\pm 0.10$ \\ 
&  &  &  & 0.386$\pm 0.$05 &  &  &  &  &  &  \\ 
$^{106}$Pd & 0.403 & 0.515 & 0.640 & 0.610$\pm 0.090$ & -0.573 & -0.648 & 
-0.722 & -0.56$\pm 0.08$ & 0.465 & 0.398$\pm 0.021$ \\ 
&  &  &  & 0.656$\pm 0.035$ &  &  &  & -0.51$\pm 0.08$ &  & 0.30$\pm 0.06$
\\ \hline
$^{108}$Cd & 0.414 & 0.531 & 0.661 & 0.540$\pm 0.011$ & -0.581 & -0.657 & 
-0.734 & -0.45$\pm 0.08$ & 0.361 & 0.34$\pm 0.09$ \\ 
&  &  &  & 0.430$\pm 0.020$ &  &  &  &  &  &  \\ 
$^{108}$Pd & 0.456 & 0.584 & 0.727 & 0.700$\pm 0.070$ & -0.610 & -0.690 & 
-0.770 & -0.58$\pm 0.04$ & 0.483 & 0.36$\pm 0.03$ \\ 
&  &  &  & 0.760$\pm 0.040$ &  &  &  & -0.51$\pm 0.06$ &  & 0.32$\pm 0.03$
\\ \hline\hline
\end{tabular}

*P.F. Mantica et al. Phy. Rev.C63, 034312 (2001). 

\pagebreak

\noindent \textbf{Table 3.} Experimental limits on half-lives $T_{1/2}^{2\nu
}(0^{+}\rightarrow 0^{+})$, theoretically calculated $M_{2\nu }$ and
corresponding $T_{1/2}^{2\nu }(0^{+}\rightarrow 0^{+})$ for 2$\nu $ $\beta
^{+}\beta ^{+},$ 2$\nu $ $\beta ^{+}$EC and 2$\nu $ ECEC decay of $^{96}$Ru, 
$^{102}$Pd, $^{106}$Cd and $^{108}$Cd nuclei. 
Half-lives are calculated using $g_{A}$= (1.261 - 1.0), respectively.

\noindent 
\begin{tabular}{cccccllc}
\hline\hline
Nuclei & Decay & \multicolumn{2}{c}{Experiment} & \multicolumn{4}{c}{Theory}
\\ 
& Mode & Ref & $T_{1/2}^{2\nu }$ (y) & Ref. & Model & $\left| M_{2\nu
}\right| $ & $T_{1/2}^{2\nu }$ (y) \\ 
&  &  &  &  &  &  &  \\ \hline
$^{96}$Ru & $\beta ^{+}\beta ^{+}$ & \cite{nor85} & \multicolumn{1}{l}{$>$
3.1$\times $10$^{16*}$} & Present & PHFB & 0.0537 & (1.378-3.485)$\times
10^{28}$ \\ 
&  &  & \multicolumn{1}{l}{} & \cite{hir94} & QRPA & 0.2510 & (6.309-15.950)$%
\times 10^{26}$ \\ 
& $\beta ^{+}$EC & \cite{nor85} & \multicolumn{1}{l}{$>$ 6.7$\times $10$%
^{16*}$} & Present & PHFB & 0.0537 & (3.599-9.100)$\times 10^{23}$ \\ 
&  &  & \multicolumn{1}{l}{} & \cite{rum98} & SU(4)$_{\sigma \tau }$ & 0.1005
& (1.028-2.598)$\times 10^{23}$ \\ 
&  &  & \multicolumn{1}{l}{} & \cite{hir94} & QRPA & 0.2694 & (1.430-3.616)$%
\times 10^{22}$ \\ 
& ECEC &  & \multicolumn{1}{l}{-} & Present & PHFB & 0.0537 & (0.644-1.628)$%
\times 10^{23}$ \\ 
&  &  & \multicolumn{1}{l}{} & \cite{rum98} & SU(4)$_{\sigma \tau }$ & 0.1005
& (1.839-4.649)$\times 10^{22}$ \\ 
&  &  & \multicolumn{1}{l}{} & \cite{hir94} & QRPA & 0.2620 & (2.705-6.840)$%
\times 10^{21}$ \\ \hline
$^{102}$Pd & $\beta ^{+}$EC &  & \multicolumn{1}{l}{} & Present & PHFB & 
0.0524 & (2.509-6.344)$\times 10^{32}$ \\ 
& ECEC &  & \multicolumn{1}{l}{} & Present & PHFB & 0.0524 & (3.783-9.565)$%
\times 10^{24}$ \\ \hline
$^{106}$Cd & $\beta ^{+}\beta ^{+}$ & \cite{dan03} & \multicolumn{1}{l}{$>$%
5.0$\times $10$^{18}$} & Present & PHFB & 0.0819 & (3.495-8.836)$\times
10^{27}$ \\ 
&  & \cite{bel99} & \multicolumn{1}{l}{$>$2.4$\times $10$^{20**}$} & \cite
{sto03} & SQRPA(l.b.) & 0.61 & (6.304-15.940)$\times 10^{25}$ \\ 
&  & \cite{bar96} & \multicolumn{1}{l}{$>$1.0$\times $10$^{19*}$} &  & 
SQRPA(s.b.) & 0.57 & (7.220-18.260)$\times 10^{25}$ \\ 
&  & \cite{dan96} & \multicolumn{1}{l}{$>$9.2$\times $10$^{17}$} & \cite
{suh01} & QRPA(AWS) & 0.722 & (4.500-11.380)$\times 10^{25}$ \\ 
&  & \cite{nor84} & \multicolumn{1}{l}{$>$2.6$\times $10$^{17*}$} &  & 
QRPA(WS) & 0.166 & (8.513-21.520)$\times 10^{26}$ \\ 
&  & \cite{win55} & \multicolumn{1}{l}{$>$6$\times 10^{16}$} & \cite{bar96}
& QRPA(WS) & 0.840 & (3.324-8.406)$\times 10^{25}$ \\ 
&  &  & \multicolumn{1}{l}{} &  & QRPA(AWS) & 0.780 & (3.856-9.749)$\times
10^{25}$ \\ 
&  &  & \multicolumn{1}{l}{} & \cite{hir94} & QRPA & 0.218 & (4.936-12.480)$%
\times 10^{26}$ \\ 
&  &  &  & \cite{sta91} & QRPA &  & 4.940$\times 10^{25}$ \\ 
& $\beta ^{+}$EC & \cite{dan03} & \multicolumn{1}{l}{$>$1.2$\times $10$^{18}$%
} & Present & PHFB & 0.0819 & (9.489-23.992)$\times 10^{22}$ \\ 
&  & \cite{bel99} & \multicolumn{1}{l}{$>$4.1$\times $10$^{20}$} & \cite
{sto03} & SQRPA(l.b.) & 0.61 & (1.712-4.328)$\times $10$^{21}$ \\ 
&  & \cite{bar96} & \multicolumn{1}{l}{$>$0.66$\times $10$^{19*}$} &  & 
SQRPA(s.b.) & 0.57 & (1.960-4.957)$\times 10^{21}$ \\ 
&  & \cite{dan96} & \multicolumn{1}{l}{$>$2.$6\times $10$^{17}$} & \cite
{suh01} & QRPA(AWS) & 0.718 & (1.236-3.124)$\times 10^{21}$ \\ 
&  & \cite{nor84} & \multicolumn{1}{l}{$>$5.7$\times $10$^{17*}$} &  & 
QRPA(WS) & 0.168 & (2.257-5.706)$\times 10^{22}$ \\ 
&  &  & \multicolumn{1}{l}{} & \cite{rum98} & SU(4)$_{\sigma \tau }$ & 0.1947
& (1.680-4.248)$\times 10^{22}$ \\ 
&  &  & \multicolumn{1}{l}{} & \cite{toi97} & RQRPA(AWS) & 0.56 & 
(2.031-5.136)$\times 10^{21}$ \\ 
&  &  & \multicolumn{1}{l}{} &  & RQRPA(WS) & 0.55 & (2.106-5.324)$\times
10^{21}$ \\ 
&  &  & \multicolumn{1}{l}{} & \cite{bar96} & QRPA(WS) & 0.84 & 
(9.027-22.820)$\times 10^{20}$ \\ 
&  &  & \multicolumn{1}{l}{} &  & QRPA(AWS) & 0.78 & (1.047-2.647)$\times
10^{21}$ \\ 
&  &  & \multicolumn{1}{l}{} & \cite{hir94} & QRPA & 0.352 & (5.141-13.000)$%
\times 10^{21}$ \\ 
&  &  &  & \cite{suh93} & QRPA(WS) & 0.493 & (2.621-6.626)$\times 10^{21}$
\\ 
&  &  &  &  &  & 0.660 & (1.462-3.697)$\times 10^{21}$ \\ \hline\hline
\end{tabular}

\pagebreak

\noindent Table 3 continued...

\noindent 
\begin{tabular}{ccclcllc}
\hline
$^{106}$Cd & ECEC & \cite{zub04} & $>$1.0$\times $10$^{18}$ & Present & PHFB
& 0.0819 & (1.293-3.270)$\times 10^{22}$ \\ 
&  & \cite{dan03} & $>$5.8$\times $10$^{17}$ & \cite{sto03} & SQRPA(l.b.) & 
0.61 & (2.333-5.899)$\times 10^{20}$ \\ 
&  & \cite{kie03} & $>$1.0$\times $10$^{18}$ &  & SQRPA(s.b.) & 0.57 & 
(2.672-6.756)$\times 10^{20}$ \\ 
&  & \cite{geo95} & $>$5.8$\times $10$^{17}$ & \cite{suh01} & QRPA(AWS) & 
0.718 & (1.684-4.258)$\times 10^{20}$ \\ 
&  &  & \multicolumn{1}{c}{} &  & QRPA(WS) & 0.168 & (3.076-7.780)$\times
10^{21}$ \\ 
&  &  & \multicolumn{1}{c}{} & \cite{civ98} & SSDH(Theo) & 0.28 & 
(1.107-2.800)$\times 10^{21}$ \\ 
&  &  & \multicolumn{1}{c}{} &  & SSDH(Exp) & 0.17 & (3.004-7.595)$\times
10^{21}$ \\ 
&  &  & \multicolumn{1}{c}{} & \cite{rum98} & SU(4)$_{\sigma \tau }$ & 0.1947
& (2.290-5.790)$\times 10^{21}$ \\ 
&  &  & \multicolumn{1}{c}{} & \cite{toi97} & RQRPA(AWS) & 0.56 & 
(2.768-6.999)$\times 10^{20}$ \\ 
&  &  & \multicolumn{1}{c}{} &  & RQRPA(WS) & 0.55 & (2.870-7.256)$\times
10^{20}$ \\ 
&  &  & \multicolumn{1}{c}{} & \cite{bar96} & QRPA(WS) & 0.84 & (1.230-3.111)%
$\times 10^{20^{{}}}$ \\ 
&  &  & \multicolumn{1}{c}{} &  & QRPA(AWS) & 0.78 & (1.427-3.608)$\times
10^{20}$ \\ 
&  &  & \multicolumn{1}{c}{} & \cite{hir94} & QRPA & 0.270 & (1.191-3.011)$%
\times 10^{21}$ \\ 
&  &  & \multicolumn{1}{c}{} & \cite{suh93} & QRPA(WS) & 0.493 & 
(3.572-9.031)$\times 10^{20}$ \\ 
&  &  & \multicolumn{1}{c}{} &  &  & 0.660 & (1.993-5.039)$\times 10^{20}$
\\ \hline
$^{108}$Cd & ECEC & \cite{dan03} & \multicolumn{1}{c}{$>$4.1$\times $10$^{17}
$} & Present & PHFB & 0.0952 & (3.939-9.959)$\times 10^{27}$ \\ 
&  & \cite{kie03} & \multicolumn{1}{c}{$>$1.0$\times $10$^{18}$} &  &  &  & 
\\ 
&  & \cite{geo95} & \multicolumn{1}{c}{$>$4.1$\times $10$^{17}$} &  &  &  & 
\\ \hline\hline
\end{tabular}

* and ** denote half-life limit for 0$\nu $ + 2$\nu $ and 0$\nu $ + 2$\nu
+0\nu $M modes respectively.

\pagebreak

\noindent \textbf{Table 4.}Effect of the variation in $\chi _{qq}$ on $%
\left\langle Q_{0}^{2}\right\rangle $, $\beta _{2}$ and NTMEs $M_{2\nu }$.

\noindent 
\begin{tabular}{cccccccccccccccc}
\hline\hline
& $\chi _{qq}$ & 0.00 & 0.20 & 0.40 & 0.60 & 0.80 & 0.90 & 0.95 & 1.00 & 1.05
& 1.10 & 1.20 & 1.30 & 1.40 & 1.50 \\ 
&  &  &  &  &  &  &  &  &  &  &  &  &  &  &  \\ 
$^{96}$Ru & $\left\langle Q_{0}^{2}\right\rangle $ & 0.0 & 0.006 & 0.214 & 
0.144 & 23.854 & 30.351 & 32.485 & 34.473 & 36.42 & 38.15 & 66.54 & 70.05 & 
73.61 & 77.48 \\ 
& $\beta _{2}$ & 0.0 & 0.046 & 0.097 & 0.098 & 0.112 & 0.140 & 0.151 & 0.161
& 0.171 & 0.180 & 0.295 & 0.317 & 0.337 & 0.348 \\ 
$^{96}$Mo & $\left\langle Q_{0}^{2}\right\rangle $ & 0.0 & 0.695 & 0.211 & 
0.477 & 22.464 & 31.82 & 37.02 & 41.73 & 45.15 & 48.15 & 61.43 & 65.44 & 
66.70 & 67.64 \\ 
& $\beta _{2}$ & 0.0 & 0.091 & 0.093 & 0.093 & 0.106 & 0.149 & 0.174 & 0.191
& 0.210 & 0.224 & 0.268 & 0.281 & 0.286 & 0.290 \\ 
& $M_{2\nu }$ & 0.168 & 0.154 & 0.152 & 0.153 & 0.093 & 0.072 & 0.067 & 0.054
& 0.036 & 0.024 & 0.076 & 0.049 & 0.056 & 0.049 \\ 
&  &  &  &  &  &  &  &  &  &  &  &  &  &  &  \\ 
$^{102}$Pd & $\left\langle Q_{0}^{2}\right\rangle $ & 0.0 & 0.252 & 0.081 & 
1.080 & 1.839 & 36.08 & 42.32 & 45.47 & 47.91 & 49.63 & 52.67 & 56.37 & 84.71
& 85.71 \\ 
& $\beta _{2}$ & 0.0 & 0.085 & 0.046 & 0.090 & 0.092 & 0.149 & 0.172 & 0.185
& 0.196 & 0.203 & 0.217 & 0.234 & 0.349 & 0.353 \\ 
$^{102}$Ru & $\left\langle Q_{0}^{2}\right\rangle $ & 0.0 & 0.028 & 0.123 & 
0.492 & 38.04 & 49.81 & 53.57 & 56.51 & 58.87 & 61.19 & 66.93 & 88.17 & 88.73
& 89.25 \\ 
& $\beta _{2}$ & 0.0 & 0.036 & 0.068 & 0.092 & 0.159 & 0.204 & 0.220 & 0.232
& 0.242 & 0.252 & 0.279 & 0.362 & 0.365 & 0.367 \\ 
& $M_{2\nu }$ & 0.178 & 0.203 & 0.208 & 0.215 & 0.135 & 0.092 & 0.072 & 0.052
& 0.039 & 0.027 & 0.021 & 0.0001 & 0.015 & 0.014 \\ 
&  &  &  &  &  &  &  &  &  &  &  &  &  &  &  \\ 
$^{106}$Cd & $\left\langle Q_{0}^{2}\right\rangle $ & 0.0 & 0.008 & 0.031 & 
0.128 & 0.510 & 32.65 & 40.83 & 47.14 & 55.89 & 62.54 & 73.53 & 83.65 & 90.63
& 91.33 \\ 
& $\beta _{2}$ & 0.0 & 0.007 & 0.003 & 0.035 & 0.073 & 0.127 & 0.152 & 0.176
& 0.211 & 0.243 & 0.299 & 0.325 & 0.344 & 0.347 \\ 
$^{106}$Pd & $\left\langle Q_{0}^{2}\right\rangle $ & 0.0 & 0.022 & 0.079 & 
0.192 & 0.897 & 39.54 & 46.88 & 52.12 & 56.13 & 59.31 & 66.23 & 73.87 & 79.84
& 92.21 \\ 
& $\beta _{2}$ & 0.0 & 0.016 & 0.042 & 0.064 & 0.094 & 0.158 & 0.183 & 0.203
& 0.216 & 0.227 & 0.254 & 0.290 & 0.322 & 0.364 \\ 
& $M_{2\nu }$ & 0.169 & 0.164 & 0.162 & 0.166 & 0.170 & 0.127 & 0.095 & 0.082
& 0.084 & 0.066 & 0.041 & 0.027 & 0.004 & 0.001 \\ 
&  &  &  &  &  &  &  &  &  &  &  &  &  &  &  \\ 
$^{108}$Cd & $\left\langle Q_{0}^{2}\right\rangle $ & 0.0 & 0.015 & 0.047 & 
0.132 & 0.488 & 35.73 & 43.87 & 53.29 & 67.98 & 76.17 & 79.26 & 80.61 & 84.18
& 94.34 \\ 
& $\beta _{2}$ & 0.0 & 0.001 & 0.013 & 0.039 & 0.076 & 0.134 & 0.161 & 0.195
& 0.258 & 0.299 & 0.311 & 0.316 & 0.326 & 0.352 \\ 
$^{108}$Pd & $\left\langle Q_{0}^{2}\right\rangle $ & 0.0 & 0.047 & 0.100 & 
0.267 & 28.21 & 44.76 & 50.99 & 55.88 & 59.63 & 62.85 & 70.17 & 76.61 & 83.14
& 86.42 \\ 
& $\beta _{2}$ & 0.0 & 0.046 & 0.051 & 0.076 & 0.125 & 0.174 & 0.196 & 0.213
& 0.225 & 0.236 & 0.267 & 0.299 & 0.334 & 0.347 \\ 
& $M_{2\nu }$ & 0.208 & 0.199 & 0.203 & 0.204 & 0.183 & 0.120 & 0.093 & 0.095
& 0.084 & 0.052 & 0.029 & 0.020 & 0.011 & 0.001 \\ \hline\hline
\end{tabular}


\begin{thebibliography}{99}
\bibitem{ros65}  S. P. Rosen and H .Primakoff, in \textit{Alpha-beta-gamma
ray spectroscopy}, ed. K. Siegbahn (1965) p.1499.

\bibitem{ver83}  J. D. Vergados, Nucl. Phys. B 218 (1983) 109.

\bibitem{ver86}  J. D. Vergados, Phys. Rep. 133 (1986) 1.

\bibitem{doi92}  M. Doi and T. Kotani, Prog. Theor. Phy. 87 (1992) 1207.

\bibitem{doi93}  M. Doi and T. Kotani, Prog. Theor. Phy. 89 (1993) 139.

\bibitem{bar95}  A. S. Barabash, \textit{Proc. Int. Workshop on Double Beta
Decay and Related Topics}, Trento, Italy, 1995, World Scientific, Singapore
(1996) 502.

\bibitem{suh98}  J. Suhonen and O. Civitarese, Phys. Rep. 300 (1998) 123.

\bibitem{kir00}  I. V. Kirpichnikov, Phys. At. Nucl. 63 (2000) 1341.

\bibitem{kla01}  H. V. Klapdor-Kleingrothaus, \textit{Sixty years of Double
Beta Decay, }World Scientific, Singapore (2001).

\bibitem{tretyak}  V. I. Tretyak, Y. G. Zdesenko, At. Data Nucl. Data Tables 
\textbf{61} (1995) 43; ibid 80 (2002) 83.

\bibitem{bar04}  A. S. Barabash, Phys. At. Nucl. 67 (2004) 438.

\bibitem{win55}  R. G. Winter, Phys. Rev. 99 (1955) 88.

\bibitem{kim83}  C. W. Kim and K. Kubodera Phys. Rev. D\textbf{\ }27 (1983)
2765.

\bibitem{aba84}  J. Abad, A. Morales, R. Nunez-Lagos and A.F. Pacheco,
Anales de Fisica A\textbf{\ }80 (1984) 15 ; J. de Phys. 45 (1984) C3-147.

\bibitem{zel81}  Ya. V. Zeldovich and M. Yu Khlopov, Pisma v ZhETF 34 (1981)
148 [JETF Lett. 34 (1981) 141].

\bibitem{era82}  R. A. Eramzhyan, G. V. Micelmacher and M. E. Voloshin,
Pisma v ZhETF 35 (1982) 530.

\bibitem{ber83}  J. Bernabeu, A. De Rujula and C. Jarlskog, Nucl. Phys. B
223 (1983) 15.

\bibitem{bal89}  S. K. Balaev, A. A. Kuliev and D. I. Salamov, Izvestiva
Akademii Nauk USSR, ser. fiz. 53 (1989) 2136 (in Russian).

\bibitem{bel82}  E. Bellotti, E. Fiorini, C. Liguori, A. Pullia, A.
Sarracino and L. Zanotti, Lett. Nuovo Cim. 33 (1982) 273.

\bibitem{nor84}  E. B. Norman and A. DeFaccio, Phys. Lett B 148\textbf{\ }%
(1984) 31.

\bibitem{nor85}  E. B. Norman Phys. Rev. C\textbf{\ }31 (1985) 1937.

\bibitem{vog86}  P. Vogel and M. R. Zirnbauer, Phys. Rev. Lett. 57 (1986)
3148.

\bibitem{bar90}  A. S. Barabash, JETP Lett. 51 (1990) 207.

\bibitem{geo95}  A. Sh. Georgadze\textit{, }F. A. Danevich, Yu. G. Zdesenko,
V. V. Kobychev, B. N. Kropivyanskii, V. N. Kuts, A. S. Nikolaiko and V. I.
Tretyak, Phys. At. Nucl 58 (1995) 1093.

\bibitem{dan96}  F. A. Danevich \textit{et al.}, Z. Phys. A 355 (1996) 433.

\bibitem{bar96}  A. S. Barabash, V.I.Umatov, R.Gurriaran, F.Hubert,
Ph.Hubert, M.Aunola,and J.Suhonen, Nucl. Phys. A 604 (1996) 115.

\bibitem{bel99}  P. Belli, R. Bernabei, A. Incicchitti, C. Arpesella, V. V.
Kobychev, O. A. Ponkratenko, V. I. Tretyak and Yu. G. Zdesenko,
Astroparticle Phys. 10 (1999) 115.

\bibitem{kie03}  H. Kiel, D. M\"{u}nstermann, K. Zuber, Nucl.Phys. A\textbf{%
\ }23 (2003) 499, arXiv: nucl-ex/0301007.

\bibitem{dan03}  F. A. Danevich, A. Sh. Georgadze, V. V. Kobychev, B. N.
Kropivyansky, A. S. Nikolaiko, O. A. Ponkratenko, V. I. Tretyak, S. Yu.
Zdesenko and Yu. G. Zdesenko, Phys. Rev. C\textbf{\ }68 (2003) 035501.

\bibitem{zub04}  K. Zuber, Eur. Phys. J. C 33 (2004) 817.

\bibitem{ito97}  Yutaka Ito, Makoto Minowa, Wataru Ootani, Keiji Nishigaki,
Yasuhiro Kishimoto, Takayuki Watanabe and Youiti Ootuka, Nucl. Inst. and
Meth. in Phys. Res. A 386, 439 (1997).

\bibitem{zub01}  K. Zuber, Phys. Lett. B\textbf{\ }519\textbf{\ } (2001) 1,
arXiv: nucl-ex/0105018.

\bibitem{Chan05}  R. Chandra, J. Singh, P.K. Rath, P.K. Raina, and J.G.
Hirsch, Eur. Phys. J. A\textbf{\ }23 (2005) 223.

\bibitem{suh01}  J. Suhonen and O. Civitarese, Phys. Lett. B\textbf{\ }497
(2001) 221.

\bibitem{shu05}  A. Shukla, P. K. Raina, R. Chandra, P. K. Rath and J. G.
Hirsch, Eur. Phys. J. A\textbf{\ }23 (2005) 235.

\bibitem{aue93}  N. Auerbach, D. C. Zheng, L. Zamick and B. A. Brown, Phys.
Lett.B\textbf{\ }304 (1993) 17; N. Auerbach, G. F. Bertsch, B. A. Brown and
L. Zhao, Nucl. Phys. A\textbf{\ }56 (1993) 190.

\bibitem{tro96}  D. Troltenier, J. P. Draayer and J. G. Hirsch, Nucl. Phys.
A \textbf{\ }601 (1996) 89.

\bibitem{bar68}  M. Baranger and K. Kumar, Nucl. Phys. A\textbf{\ }110
(1968) 490.

\bibitem{dix02}  B. M. Dixit, P. K. Rath and P. K. Raina Phys. Rev. C\textbf{%
\ }65 (2002) 034311, Phys. Rev. C 67 (2003) 059901(E).

\bibitem{civ93}  O. Civitarese and J. Suhonen, Phys. Rev. C\textbf{\ }47
(1993) 2410.

\bibitem{boh98}  A. Bohr and B, R. Mottelson, Nuclear Structure Vol. I
(World Scientific, Singapore, 1998).

\bibitem{cas94}  O. Casta\~{n}os, J. G. Hirsch, O. Civitarese and P. O.
Hess, Nucl. Phys. A 571 (1994) 276.

\bibitem{hir95}  J. G. Hirsch, O. Casta\~{n}os, P. O. Hess and O.
Civitarese, Phys. Rev. C\textbf{\ }51 (1995) 2252.

\bibitem{cer99}  V. E.Ceron and J.G.Hirsch, Phys. Lett.B 471 (1999) 1.

\bibitem{oni66}  N. Onishi and S. Yoshida, Nucl. Phys. A\textbf{\ }260
(1966) 226.

\bibitem{ver71}  J. D. Vergados and T.T.S. Kuo, Phys. Lett. B 35 (1971) 93.

\bibitem{kho82}  S. K. Khosa, P. N. Tripathi and S. K. Sharma, Phys. Lett. B
119 (1982) 257; P. N. Tripathi, S. K. Sharma and S. K. Khosa, Phys. Rev. C
29 (1984) 1951; S. K. Sharma, P. N. Tripathi and S. K. Khosa, Phys. Rev. C
38 (1988) 2935.

\bibitem{hee69}  G. M. Heestand, R. R. Borchers, B. Herskind, L. Grodzins,
R. Kalish and D. E. Murnick, Nucl. Phys. A\textbf{\ }133 (1969) 310.

\bibitem{gre66}  W. Greiner, Nucl. Phys. 80 (1966) 417.

\bibitem{ari81}  A. Arima, Nucl. Phys. A\textbf{\ }354 (1981) 19.

\bibitem{sak84}  M. Sakai At. Data and Nucl. Data Tables 31 (1984) 400.

\bibitem{ram87}  S. Raman, C. H. Malarkey, W. T. Milner, C. W. Nestor, JR.
and P. H. Stelson, At. Data and Nucl. Data Tables 36 (1987) 1.

\bibitem{rag89}  P. Raghavan, Atomic Data and Nuclear Data Table 42 (1989)
189.

\bibitem{hir94}  M. Hirsch, M. Muto, T. Oda, H. V. Klapdor- Kleingrothaus,
Z. Phys. A\textbf{\ }347 (1994) 151.

\bibitem{rum98}  O. A. Rumyantsev and M. G. Urin, Phys. Lett. B\textbf{\ }%
443 (1998) 51.

\bibitem{sta91}  A. Staudt, K. Muto and H. V. Klapdor- Kleingrothaus Phys.
Lett. B\textbf{\ }268 (1991) 312.

\bibitem{suh93}  J. Suhonen, Phys. Rev. C\textbf{\ }48 (1993) 574.

\bibitem{toi97}  J. Toivanen and J. Suhonen, Phys. Rev. C 55 (1997) 2314.

\bibitem{civ98}  O. Civitarese and J. Suhonen, Phys. Rev. C\textbf{\ }58
(1998) 1535.

\bibitem{sto03}  S. Stoica, and H.V. Klapdor-Kleingrothaus Eur. Phys. J. A
17 (2003) 529.

\pagebreak 
\end{thebibliography}
\end{document}